\documentclass[UKenglish]{book}
\usepackage[dvips]{graphicx}
\usepackage[sectionbib,sort&compress]{natbib}
\usepackage{chapterbib}
\usepackage{wileyvch}
\usepackage[breaklinks]{hyperref}
\usepackage{microtype}
\usepackage{color}

\usepackage{oldgerm,amsmath,pifont,amsfonts,amssymb,latexsym}
\usepackage{float,xspace,calc,theorem,ifthen}
\usepackage{enumerate,psboxit}
\usepackage{hyperref}
\usepackage{changebar}

\newcommand{\pard}[2]{\frac{\partial #1}{\partial #2}}

\newcommand{\be}{\begin{equation}}
\newcommand{\ee}{\end{equation}}
\newcommand{\ba}{\begin{eqnarray*}}
\newcommand{\ea}{\end{eqnarray*}}
\newcommand{\bna}{\begin{eqnarray}}
\newcommand{\ena}{\end{eqnarray}}

\newcommand{\mpaa}{\begin{minipage}[t]{7.5cm}}
\newcommand{\mpea}{\end{minipage}}
\newcommand{\bc}{\begin{center}}
\newcommand{\ec}{\end{center}}

\author{Rainer Klages, Aleksei V.~Chechkin and Peter Dieterich}
\title{Anomalous Fluctuation Relations}

\begin{document}

\chapterauthor{Rainer Klages, Aleksei V.~Chechkin and Peter Dieterich}
\chapter{Anomalous fluctuation relations}

{\setlength{\baselineskip}{1\baselineskip} \small
\textbf{\textsf{Abstract.}} We study Fluctuation Relations (FRs) for
dynamics that are anomalous, in the sense that the diffusive properties
strongly deviate from the ones of standard Brownian motion. We first
briefly review the concept of transient work FRs for stochastic
dynamics modeled by the ordinary Langevin equation. We then introduce
three generic types of dynamics generating anomalous diffusion: L\'evy
flights, long-time correlated Gaussian stochastic processes and
time-fractional kinetics. By combining Langevin and kinetic approaches
we calculate the work probability distributions in the simple
nonequilibrium situation of a particle subject to a constant
force. This allows us to check the transient FR for anomalous
dynamics. We find a new form of FRs, which is intimately related to
the validity of fluctuation-dissipation relations. Analogous results
are obtained for a particle in a harmonic potential dragged by a
constant force. We argue that these findings are important for
understanding fluctuations in experimentally accessible systems. As an
example, we discuss the anomalous dynamics of biological cell
migration both in equilibrium and in nonequilibrium under chemical
gradients.  \par}

\section{Introduction}

With {\em Fluctuation Relations} (FRs) we denote a set of symmetry
relations describing large-deviation properties (see the Chapter by
Touchette and Harris in this book) of the probability distribution
functions (PDFs) of statistical physical observables far from
equilibrium. First forms of one subset of them, often referred to as
{\em Fluctuation Theorems}, emerged from generalizing
fluctuation-dissipation relations to nonlinear stochastic processes
\cite{BoKu81a,BoKu81b}. They were then discovered as generalizations
of the Second Law of Thermodynamics for thermostated dynamical
systems, i.e., systems interacting with thermal reservoirs, in
nonequilibrium steady states \cite{ECM93,EvSe94,GaCo95a,GaCo95b}; see
the Chapters by Reid et al.\ and by Rondoni and Jepps for this
deterministic approach. Another subset, so-called {\em Work
Relations}, generalize a relation between work and free energy, known
from equilibrium thermodynamics, to nonequilibrium situations
\cite{Jar97a,Jar97b}; see the Chapters by Alemany et al.\ and by
Spinney and Ford for this line of research. These two fundamental
classes were later on amended and generalized by a variety of other
FRs from which they can partially be derived as special cases
\cite{Croo99,HaSa01,Sei05,SaUe10}, as has already been discussed
starting from the Chapters by Spinney and Ford up to the one by Gaspard
in this book.  Research performed over the past ten years has shown
that FRs hold for a great variety of systems thus featuring one of the
rare statistical physical principles that is valid even very far from
equilibrium: see summaries in
\cite{HaSch07,Kurch07,Seif08,Jarz08,BoWyJa10,vdB10} for stochastic
processes, \cite{Gall98,EvSe02,Kla06,RoMM07,JeRo10,JaPiRB11} for
deterministic dynamics and \cite{EsHM09,CaHaTa11} for quantum
systems. Many of these relations have meanwhile been verified in
experiments on small systems, i.e., systems on molecular scales
featuring only a limited number of relevant degrees of freedom
\cite{WSM+02,Rit03,BLR05,CJP10,TSUMS10,AlRiRi11}, cf.\ the Chapters by
Ciliberto et al., Alemany et al., and Sagawa and Ueda.

The term {\em anomalous} in the title of this chapter refers to {\em
anomalous dynamics}, which are loosely speaking processes that do not
obey the laws of conventional statistical physics and thermodynamics
\cite{BoGe90,MeKl00,KlSo11}; see, e.g., the Chapter by Zhang et al.\
for anomalous deviations from Fourier's Law of heat conduction in
small systems. Paradigmatic examples are diffusion processes where the
long-time mean square displacement does not grow linearly in time:
That is, $\langle x^2\rangle \sim t^\alpha$, where the angular
brackets denote an ensemble average, does not increase with $\alpha=1$
as expected for Brownian motion but either {\em subdiffusively} with
$\alpha<1$ or {\em superdiffusively} with $\alpha>1$
\cite{SZK93,KSZ96,SKB02}. After pioneering work on amorphous
semiconductors \cite{SM75}, anomalous transport phenomena have more
recently been observed in a wide variety of complex systems, such as
plasmas \cite{Bale05}, nanopores \cite{Kuk96}, epidemic spreading
\cite{BHG06}, biological cell migration \cite{DKPS08} and glassy
materials \cite{BBW08}, to mention a few \cite{MeKl04,KRS08}. This
raises the question to which extent conventional FRs are valid for
anomalous dynamics. Theoretical results for generalized Langevin
equations \cite{BeCo04,OhOh07,MaDh07,CCC08}, L\'evy flights
\cite{ToCo07,ToCo09} and Continuous-Time Random Walk models
\cite{EsLi08} as well as computer simulations for glassy dynamics
\cite{Sell09} showed both validity and violations of the various types
of conventional FRs referred to above, depending on the specific type
of anomalous dynamics considered and the nonequilibrium conditions
that have been applied \cite{ChKl09}.

The purpose of this chapter is to outline how the two different fields
of FRs and anomalous dynamics can be cross-linked in order to explore
to which extent conventional forms of FRs are valid for anomalous
dynamics. With the term {\em Anomalous Fluctuation Relations} we refer
to deviations from conventional forms of FRs as they have been
discussed in the previous chapters, which are due to anomalous
dynamics. Here we focus on generic types of stochastic anomalous
dynamics by only checking {\em Transient Fluctuation Relations}
(TFRs), which describe the approach from a given initial distribution
towards a (non)equilibrium steady state. Section \ref{sec:motiv}
motivates the latter type of FRs by introducing simple scaling
relations, as they are partially used later on in this chapter. As a
warm-up, we then first derive the conventional TFR for the trivial
case of Brownian motion of a particle moving under a constant external
force modeled by standard Langevin dynamics. Section \ref{sec:afr}
introduces three generic types of stochastic anomalous dynamics:
long-time correlated Gaussian stochastic processes, L\'evy flights and
time-fractional kinetics. We check these three stochastic models for
the existence of conventional TFRs under the simple nonequilibrium
condition of a constant external force. Section \ref{sec:cell}
introduces a system exhibiting anomalous dynamics that is
experimentally accessible, which is biological cell migration. We then
outline how an anomalous transient fluctuation relation might be
verified for cells migrating under chemical gradients.  We summarize
our results in Section \ref{sec:suco} by highlighting an intimate
connection between the validity of conventional TFRs and the validity
of fluctuation-dissipation relations.

\section{Transient fluctuation relations}\label{sec:motiv}

\subsection{Motivation}

\begin{figure}[t]  
\centerline{\includegraphics[width=5cm]{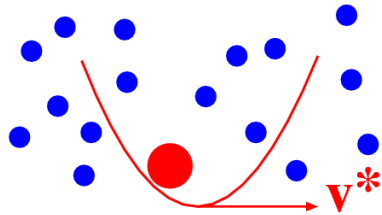}}
\caption{Sketch of a colloidal particle confined within a harmonic
trap that is dragged through water with a constant velocity $v^*$, cf.\ the 
experiment by Wang et al.\ \cite{WSM+02}.}
\label{fig:patrap}
\end{figure}

Consider a particle system evolving from some initial state at time
$t=0$ into a nonequilibrium steady state for $t\to\infty$. A famous
example that has been investigated experimentally \cite{WSM+02}, cf.\
also the Chapter by Alemany et al.\, is a colloidal particle immersed
into water and confined by an optical harmonic trap, see
Fig.~\ref{fig:patrap}. The trap is first at rest but then dragged
through water with a constant velocity $v^*$. Another paradigmatic
example, whose nonequilibrium fluctuations have been much studied by
molecular dynamics computer simulations \cite{ECM93}, is an
interacting many-particle fluid under a shear force, which starts in
thermal equilibrium by evolving into a nonequilibrium steady state
\cite{EvSe94}.

The key for obtaining FRs in such systems is to obtain the PDF
$\rho(\xi_t)$ of suitably defined dimensionless entropy
production $\xi_t$ over trajectory segments of time length $t$. The
goal is to quantify the asymmetry between positive and negative
entropy production in $\rho(\xi_t)$ for different times $t$ since, as
we will demonstrate in a moment, this relation is intimately related
to the Second Law of Thermodynamics. For a very large class of
systems, and under rather general conditions, it was shown that the
following equation holds
\cite{EvSe02,RoMM07,HaSch07,Kurch07,JeRo10}:
\be  
\ln\frac{\rho(\xi_t)}{\rho(-\xi_t)}=\xi_t \label{eq:tfr}\quad .
\ee
Given that here we consider the transient evolution of a system from
an initial into a steady state, this formula became known as the {\em
transient fluctuation relation} (TFR). The left hand side we may call
the fluctuation ratio. Relations exhibiting this functional form
have first been proposed in the seminal work by Evans, Cohen and
Morriss \cite{ECM93}, although in the different situation of
considering nonequilibrium steady states. Such a steady state relation
was proved a few years later on by Gallavotti and Cohen for
deterministic dynamical systems, based on the so-called chaotic
hypothesis \cite{GaCo95a,GaCo95b}. The idea to consider such relations
for transient dynamics was first put forward by Evans and Searles
\cite{EvSe94}.

\begin{figure}[t]
\centerline{\includegraphics[width=6cm,angle=-90]{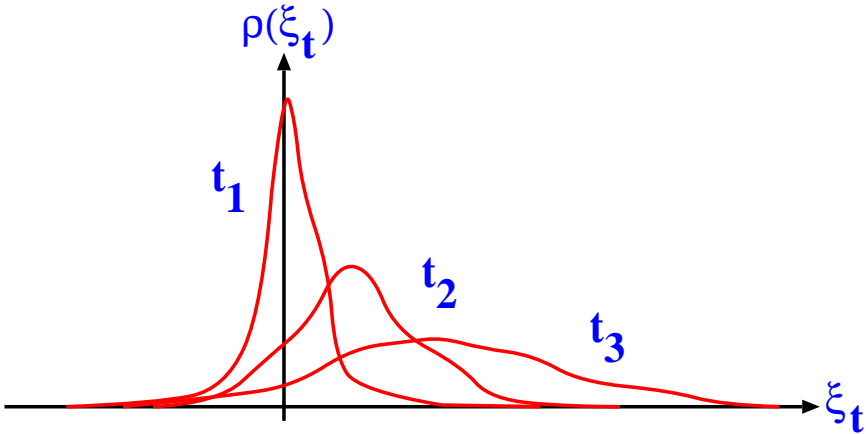}}
\caption{Illustration of the dynamics of the probability density function for
entropy production $\rho(\xi_t)$ for different times $t_1<t_2<t_3$.}
\label{fig:eppdf} 
\end{figure}

Fig.~\ref{fig:eppdf} displays the temporal evolution of the PDF
for entropy production in such a situation and may
be compared to Fig.~12 in the chapter of (Ciliberto) for analogous
results extracted from experimental measurements. The asymmetry of the
evolving distribution, formalized by the fluctuation relation
Eq.~(\ref{eq:tfr}), is in line with the Second Law of Thermodynamics. This
easily follows from Eq.~(\ref{eq:tfr}) by noting that
\be
\rho(\xi_t)=\rho(-\xi_t)\exp(\xi_t)\ge\rho(-\xi_t)\;,
\ee
where $\xi_t$ is taken to be positive or zero. Integration from zero
to infinity over both sides of this inequality after multiplication
with $\xi_t$ and defining the ensemble average over the given PDF as
$\langle\ldots\rangle=\int_{-\infty}^{\infty}d\xi_t\;\rho(\xi_t)\ldots$
yields
\be
\langle\xi_t\rangle\ge0\:.\label{eq:frsl}
\ee

\subsection{Scaling}

By using FRs one is typially interested in assessing large deviation
properties of the PDF of entropy production. That is, one wishes to
sample the tails of the distributions for large times, and not so much
the short-time dynamics, or the centre of the distribution. For this
purpose it is useful to introduce suitably scaled variables that
enable us to eliminate the drift associated with the positive average
entropy production Eq.~(\ref{eq:frsl}). A first option is to look at
the PDF $\rho(\tilde{\xi}_t)$ of the scaled variable \cite{ToCo09}
\begin{figure}[t]
\centerline{\includegraphics[width=6cm,angle=-90]{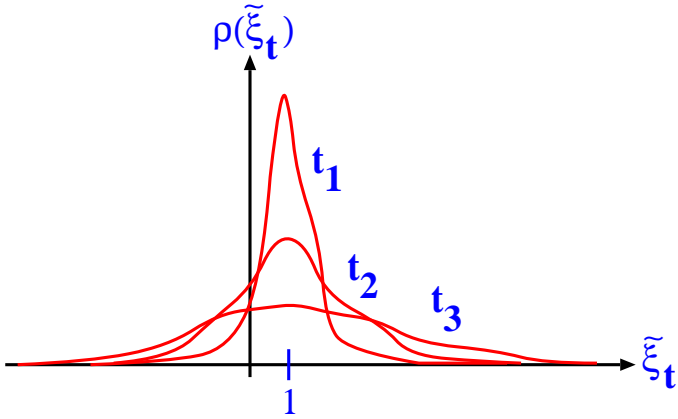}}
\caption{Illustration of the dynamics of the probability density
function for entropy production $\rho(\tilde{\xi}_t)$ for different
times $t_1<t_2<t_3$ by using the scaled variable
Eq.~(\ref{eq:scale1}).}  \label{fig:scale1} 
\end{figure} 
\be
\tilde{\xi}_t=\frac{\xi_t}{\langle\xi_t\rangle}\:, \label{eq:scale1} 
\ee 
as illustrated in Fig.~\ref{fig:scale1}. By definition, the PDF is now
centred at $\langle\tilde{\xi}_t\rangle =1$, hence we have eliminated
any contributions to the left hand side of Eq.~(\ref{eq:tfr})
that comes from the drift, by purely focusing on the asymmetric
shape of the distribution.

Another way of scaling was used by Gallavotti and Cohen
\cite{GaCo95a,GaCo95b} by employing the scaled time average
\begin{figure}[t]
\centerline{\includegraphics[width=6cm,angle=-90]{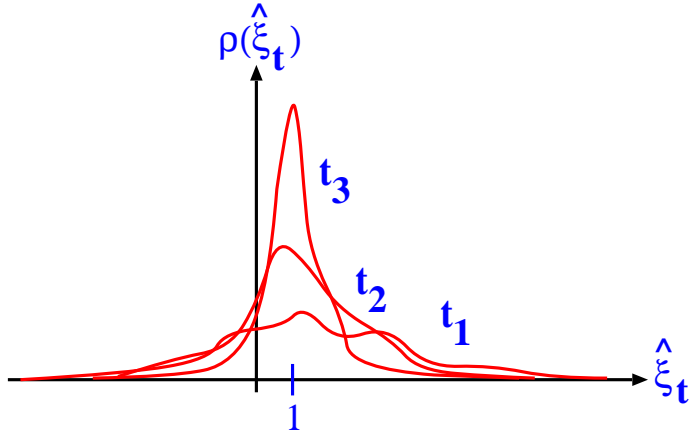}}
\caption{Illustration of the dynamics of the probability density function for
entropy production $\rho(\hat{\xi}_t)$ for different times
$t_1<t_2<t_3$ by using the scaled time average Eq.~(\ref{eq:scale2}).}
\label{fig:scale2}
\end{figure}
\be
\hat{\xi}_t=\frac{\xi_t}{t\langle\xi_t\rangle}\:, \label{eq:scale2}
\ee
yielding the PDF for entropy production displayed in
Fig.~\ref{fig:scale2}. With this scaling, and for ergodic systems, clearly
\be
\rho(\hat{\xi}_t)\to\delta(1-\hat{\xi}_t)\quad (t\to\infty)
\ee
with
\be
\frac{\xi_t}{t}\to\langle\xi_t\rangle\ge0\quad (t\to\infty)\quad .
\ee
thus illustrating the relation between FRs and the Second Law again.

\subsection{Transient fluctuation relation for ordinary Langevin dynamics}\label{sec:tfrole}

As a preparation for what follows, we may first check the TFR for the
ordinary overdamped {\em Langevin equation} \cite{KTH92}
\be
\dot{x}=F+\zeta(t)\quad , \label{eq:ole}
\ee
with a constant external force given by  $F$ and Gaussian white noise
$\zeta(t)$. Note that for sake of simplicity, here we set all the
other constants that are not relevant within this specific context
equal to one. For Langevin dynamics with a constant
force the entropy production $\xi_t$ defined by the heat,
or equivalently the dissipative work, is simply equal to the mechanical work
\cite{vZCo03b}
\be
W_t=Fx(t)\quad .
\ee
It follows that the PDF for entropy production, which
here is identical to the one for the mechanical work, is trivially
related to the PDF of the position $x$ of the Langevin
particle via
\be
\rho(W_t)=F^{-1}\varrho(x,t)\quad . \label{eq:whscal}
\ee
This is very convenient, since it implies that all that remains to be
done in order to check the TFR Eq.~(\ref{eq:tfr}) is to solve the
Fokker-Planck equation for the position PDF $\varrho(x,t)$ for a given
initial condition. Here and in the following, we choose $x(0)=0$,
i.e., in terms of position PDFs we start with a delta-distribution at
$x=0$. Note that for ordinary Langevin dynamics in a given
potential, typically the equilibrium density is taken as the initial
density \cite{vZCo03,vZCo03b}. However, since in the following we will
consider dynamics that may not exhibit a simple equilibrium state,
without loss of generality here we make a different choice.

For the ordinary Langevin dynamics Eq.~(\ref{eq:ole}) modeling a
linear Gaussian stochastic process, the position PDF is Gaussian
exhibiting normal diffusion \cite{KTH92,Risk}, cf.\ also the Chapter
by Ford and Spinney,
\be
\varrho(x,t)=\frac{1}{\sqrt{2\pi\sigma_{x,0}^2}}\exp\left(-\frac{(x-\langle x\rangle )^2}{2\sigma_{x,0}^2}\right)\quad . \label{eq:gauss}
\ee
With the subscript zero we denote ensemble averages 
in case of zero external field. By using the PDF-scaling
Eq.~(\ref{eq:whscal}) and plugging this result into the TFR
Eq.~(\ref{eq:tfr}), we easily derive that the TFR for the work $W_t$
holds if
\be
\langle W_t\rangle =\frac{\sigma_{W_t,0}^2}{2} \quad , \label{eq:fdr1}
\ee
which is nothing else than an example of the {\em
Fluctuation-Dissipation Relation of the first kind} (FDR1)
\cite{Kubo66,KTH92}, cf.\ also the Chapter by Gradenigo et al. We thus arrive at
the seemingly trivial but nevertheless important result that for this
simple Gaussian stochastic process, the validity of FDR1
Eq.~(\ref{eq:fdr1}) implies the validity of the work TFR
Eq.~(\ref{eq:tfr}). For a full analysis of FRs of ordinary Langevin
dynamics we refer to van Zon and Cohen Refs.~\cite{vZCo03,vZCo03b}.

Probably inspired by the experiment of Ref.~\cite{WSM+02}, typically
Langevin dynamics in a harmonic potential moving with a constant
velocity has been studied in the literature
\cite{ZBC05,MaDh07,OhOh07,CCC08}, cf.\ Fig.~\ref{fig:patrap}. Note
that in this slightly more complicated case the (total) work is not
equal to the heat \cite{vZCo03b}. While for the work one recovers the
TFR in its conventional form Eq.~(\ref{eq:tfr}) in analogy to the
calculation above, surprisingly the TFR for heat looks different for
large enough fluctuations. This is due to the system being affected by
the singularity of the harmonic potential, as has nicely been
elucidated by van Zon and Cohen \cite{vZCo03}. A similar effect has
been reported by Harris et al.\ for a different type of stochastic
dynamics, the asymmetric zero-range process \cite{HRS06}. For
deterministic dynamics involving Nos\'e-Hoover thermostats analogous
consequences for the validity of the Gallavotti-Cohen FR have been
discussed in Ref.~\cite{ESR05}. See Ref.~\cite{HaSch07} for a brief
review about the general mechanism underlying this type of violation
of conventional forms of TFRs.

In the following we check for yet another source of deviations from
the conventional TFR Eq.~(\ref{eq:tfr}) than the one induced by
singular potentials. We explore the validity of work TFRs if one makes
the underlying microscopic dynamics more complicated by modeling
dynamical correlations or using non-Gaussian PDFs. In order to
illustrate the main ideas along these lines it suffices to consider a
nonequilibrium situation simply generated by a constant external
force.

\section{Transient work fluctuation relations for anomalous dynamics}\label{sec:afr}

Our goal is to check the TFR Eq.~(\ref{eq:tfr}) for three generic
types of stochastic processes modeling anomalous diffusion
\cite{KRS08}: (1) {\em Gaussian stochastic processes},
(2) {\em L\'evy flights}, and (3) {\em time-fractional kinetics}. All
these dynamics we model by generalized Langevin equations. This
section reports results from Ref.~\cite{ChKl09}, which may be
consulted for further details.

\subsection{Gaussian stochastic processes}\label{sec:gsp}

The first type we consider are Gaussian stochastic processes defined by the
overdamped generalized Langevin equation
\be  
\int_0^t dt'\dot{x}(t')\gamma(t-t')=F+\zeta(t) \label{eq:gle}
\ee
with Gaussian noise $\zeta(t)$ and friction that is modeled with a
memory kernel $\gamma(t)$. By using this equation a stochastic process
can be defined that exhibits normal statistics but with anomalous
memory properties in form of non-Markovian long-time correlated
Gaussian noise.  Equations of this type can be traced back at least to
work by Mori and Kubo around 1965 (see \cite{Kubo66} and further
references therein). They form a class of standard models generating
anomalous diffusion that has been widely investigated, see, e.g.,
Refs.~\cite{KTH92,PWM96,Lutz01}. FRs for this type of dynamics have
more recently been analyzed in
Refs.~\cite{BeCo04,OhOh07,MaDh07,CCC08}. Examples of applications for
this type of stochastic modeling are given by generalized elastic
models \cite{TCK10}, polymer dynamics \cite{Panja10} and biological
cell migration \cite{DKPS08}.

We now split this class into two specific cases.

\subsubsection{Correlated internal Gaussian noise}

The first case corresponds to {\em internal} Gaussian noise, in the
sense that we require the system to exhibit the {\em
Fluctuation-Dissipation Relation of the second kind} (FDR2)
\cite{Kubo66,KTH92}
\be
\langle \zeta(t)\zeta(t')\rangle\sim \gamma(t-t')\:,
\ee
again by neglecting all constants that are not relevant for the main
point we wish to make here. We now consider the specific case that
both the noise and the friction are correlated by a simple power law,
\be
\gamma(t)\sim t^{-\beta}\;,\;0<\beta<1\:.
\ee
Because of the linearity of the generalized Langevin equation
(\ref{eq:gle}) the position PDF must be the Gaussian
Eq.(\ref{eq:gauss}), and by the scaling of Eq.~(\ref{eq:whscal}) we have
$\rho(W_t)\sim\varrho(x,t)$. It thus remains to solve
Eq.~(\ref{eq:gle}) for mean and variance, which can be done in Laplace
space \cite{ChKl09} yielding {\em subdiffusion},
\be
\sigma_{x,F}^2\sim t^{\beta}\:,
\ee
by preserving the FDR1 Eq.~(\ref{eq:fdr1}). Here and in the
following we denote ensemble averages in case of a non-zero external
field with the subscript $F$. For Gaussian stochastic processes we
have seen in Section~\ref{sec:tfrole} that the conventional work TFR
follows from FDR1. Hence, for the above power-law correlated internal
Gaussian noise we recover the conventional work TFR
Eq.~(\ref{eq:tfr}).

\subsubsection{Correlated external Gaussian noise}\label{sec:cegn}

As a second case, we consider the overdamped generalized Langevin
equation
\be
\dot{x}=F+\zeta(t)\label{eq:gle2}\:,
\ee
which represents a special case of Eq.~(\ref{eq:gle}) with a
memory kernel modeled by a delta-function. Again we use correlated
Gaussian noise defined by the power law
\be
\langle\zeta(t)\zeta(t')\rangle\sim |t-t'|^{-\beta}\;,\;0<\beta<1\:,
\ee
which one may call {\em external}, because in this case we do not
postulate the existence of FDR2. The position PDF is again Gaussian,
and as before $\rho(W_t)\sim\varrho(x,t)$. However, by solving the
Langevin equation along the same lines as in the previous case, here
one obtains {\em superdiffusion} by breaking FDR1,
\be
\langle W_t\rangle\sim t\quad , \quad \sigma_{W_t,F}^2\sim t^{2-\beta}\:.
\ee
Calculating the fluctuation ratio, i.e., the left hand side of
Eq.~(\ref{eq:tfr}), from these results yields
the {\em anomalous work TFR}
\be
\ln\frac{\rho(W_t)}{\rho(-W_t)}=C_{\beta} t^{\beta-1}W_t \quad 0<\beta<1\:,\label{eq:afr}
\ee
where $C_{\beta}$ is a constant that depends on physical parameters
\cite{ChKl09}. Comparing this equation with the conventional form of
the TFR Eq.~(\ref{eq:tfr}) one observes that the fluctuation ratio is
still linear in $W_t$ thus exhibiting the exponential large deviation
form \cite{ToCo09}, cf.\ the Chapter by Touchette and Harris. However,
there are two important deviations: (1) the slope of the fluctuation
ratio as a function of $W_t$ is not equal to one anymore, and in
particular (2) it decreases with time. We may thus classify
Eq.~(\ref{eq:afr}) as a {\em weak violation of the conventional TFR}.

\begin{figure}[t]  
\centerline{\includegraphics[width=8cm]{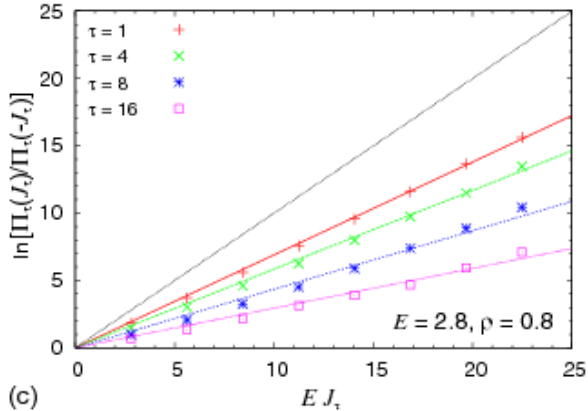}}
\caption{The fluctuation ratio 
$\ln(\Pi_{\tau}(J_{\tau})/\Pi_{\tau}(J_{\tau}))$ for the entropy
production $W_{\tau}=EJ_{\tau}$ with particle current $J_{\tau}$ and
field strength $E$ for particle density $\rho$ at different times
$\tau$. The full line, with slope one, displays the result of the
conventional FR Eq.~(\ref{eq:tfr}) in a nonequilibrium steady state.
The figure is from Ref.~\cite{Sell09}.}
\label{fig:sellafr}
\end{figure}

We remark that for driven glassy systems FRs have already been
obtained displaying slopes that are not equal to one. Within this
context it has been suggested to capture these deviations from one by
introducing the concept of an `effective temperature'
\cite{Sell98,ZRA05,ZBC05}. As far as the time dependence of the
coefficient is concerned, such behavior has recently been observed in
computer simulations of a paradigmatic two dimensional lattice gas
model generating glassy dynamics \cite{Sell09}. Fig.~\ref{fig:sellafr}
shows the fluctuation ratio as a function of the entropy production at
different times $\tau$ as extracted from computer simulations of this
model, where the PDF has first been relaxed into a nonequilibrium
steady state. It is clearly seen that the slope decreases with time,
which is in line with the prediction of the anomalous TFR
Eq.~(\ref{eq:afr}). However, to which extent the nonequilibrium
dynamics of this lattice gas model can be mapped onto the generalized
Langevin equation Eq.~(\ref{eq:gle2}) is an open question.

\subsection{L\'evy flights}

A second fundamental type of anomalous dynamics can as well be defined
by the overdamped Langevin equation (\ref{eq:gle2}). However, this
time we choose white {\em L\'evy noise}, that is, the random variable
$\zeta$ is distributed according to the PDF
\be
\chi(\zeta)\sim
\zeta^{-1-\alpha}\quad (\zeta\to\infty)\:,\:0\le\alpha<2\:.\label{eq:levy}
\ee
In general, the full L\'evy stable PDF is defined by its
characteristic function. In this case we are thus dealing with
Markovian stochastic processes that are not Gaussian distributed
generating so-called {\em L\'evy flights}, which are due to the heavy
tails of the underlying PDF. An introduction to the theory of L\'evy
flights can be found in Chapter~5 of Ref.~\cite{KRS08}; the
rigorous mathematical theory is presented in, e.g.,
Ref.~\cite{SaTa94}. L\'evy flights define one of the most
paradigmatic models of anomalous dynamics with wide applications, for
example, in fluid dynamics
\cite{SWS93}, in the foraging of biological organisms
\cite{VLRS11} and in glassy optical material
\cite{BBW08}, to highlight only a few cases.

It can be shown that the position PDF $\varrho(x,t)$ characterizing
the process defined by Eqs.~(\ref{eq:gle2}),(\ref{eq:levy}) obeys the
{\em space-fractional} Fokker-Planck equation
\be
\pard{\varrho}{t}=-F\pard{\varrho}{x}+\pard{^{\alpha}\varrho}{|x|^{\alpha}}\:,\label{eq:sffpe}
\ee
where the last term is given by the {\em Riesz fractional derivative},
which in real space is a complicated integro-differential operator. It
is thus more convenient to represent this derivative by its Fourier
transform, which takes the simple expression
\be
{\cal F}\left\{\partial^{\alpha}\varrho/\partial|x|^{\alpha}\right\}=-|k|^{\alpha}{\cal F}\left\{\varrho\right\}\;.
\ee
Fractional derivatives provide generalizations of ordinary derivatives
by reproducing them in case of integer values of the derivative
parameter. Being defined by power law memory kernels, they have proven
to be extremely useful in order to mathematically model anomalous
dynamics. The well-developed discipline of {\em fractional calculus}
rigorously explores the properties of these mathematical objects; for
introductions to fractional derivatives see, e.g.,
Refs.~\cite{Podl99,MeKl00,KRS08,KlSo11}. A systematic and
comprehensive mathematical exposition of fractional calculus is given
in Ref.~\cite{SKM93}. After solving Eq.~(\ref{eq:sffpe}) in Fourier
space, the resulting position PDF needs to be converted into the work
PDF by using Eq.~(\ref{eq:whscal}). In this case it is sensible to
apply the scaling Eq.~(\ref{eq:scale1}) \cite{ToCo09}, which here
yields the scaled variable $\tilde{W}_t=W_t/F^2t$. Expressing the work
PDF in this variable and using the asymptotics of the L\'evy stable
PDF Eq.~(\ref{eq:levy}), we arrive at the asymptotic TFR for L\'evy
flights
\be
\lim_{\tilde{W}_t\to\pm\infty}\frac{\rho(\tilde{W}_t)}{\rho(-\tilde{W}_t)}=1\: . \label{eq:tfrlevy}
\ee
This result has first been reported by Touchette and Cohen in
Ref.~\cite{ToCo07} by using a different technique for the different
situation of a harmonic potential dragged with a constant
velocity. Note that for $\alpha=2$ in the above model we recover the
conventional TFR Eq.~(\ref{eq:tfr}). For $0<\alpha<2$, however, we
obtain the surprising result that asymptotically large positive and
negative fluctuations of the scaled work are equally probable for
L\'evy flights. The underlying work PDF is nevertheless still
generically asymmetric. Note that the fluctuation ratio
Eq.~(\ref{eq:tfrlevy}) does not display the exponential large
deviation form, hence one may denote this as a {\em strong violation
of the conventional TFR}.

\subsection{Time-fractional kinetics}

The third and final fundamental type of stochastic anomalous dynamics
that we consider here can be modeled by the so-called {\em
subordinated} Langevin equation \cite{Fog94,BaFr05}
\be
\frac{dx(u)}{du}=F+\zeta(u)\quad,\quad\frac{dt(u)}{du}=\tau(u)\label{eq:sle}
\ee
with Gaussian white noise $\zeta(u)$ and white L\'evy noise
$\tau(u)>0$ with $0<\alpha<1$. It can be shown that subordinated
Langevin dynamics is intimately related to {\em Continuous Time Random
Walk Theory}, which provides a generalization of ordinary random walk
theory by generating non-trivial jump dynamics. The latter approach
has in turn been used, e.g., to understand measurements of anomalous photo
currents in copy machines \cite{SM75}, microsphere diffusion in the
cell membrane \cite{MeKl04}, translocations of biomolecules through
membrane pores \cite{MeKl03} and even dynamics of prices in financial markets
\cite{Sca06}. It was demonstrated that this Langevin description leads to the
time-fractional Fokker-Planck equation \cite{Fog94,BaFr05}
\be
\pard{\varrho}{t}=\pard{^{1-\alpha}}{t^{1-\alpha}}\left[-\pard{F\varrho}{x}+\pard{^2\varrho}{x^2}\right] \label{eq:fpesle}
\ee
for $0<\alpha<1$ with {\em Riemann-Liouville fractional derivative} on
the right semi-axis
\be
\frac{\partial^{\delta} \varrho}{\partial t^{\delta}} = 
\frac{\partial}{\partial t}\left[\frac{1}{\Gamma (1-\delta)} \int_0^t dt^{\prime }\frac{\varrho(t^{\prime})}{(t-t^{\prime })^{\delta}}\right] \label{eq:rlfd}
\ee
for $0<\delta<1$. This equation obeys a (generalized) Einstein
relation for friction and diffusion coefficients (which here are both
set to unity, for sake of simplicity). From Eq.~(\ref{eq:fpesle}),
equations for the first and second moments can be derived and then
solved in Laplace space.  The second moment in the absence of an
external force yields {\em subdiffusion},
\be
\sigma_{x,0}^2\sim t^{\alpha}\:.
\ee
A calculation of the current $\langle x\rangle$ shows that the FDR1
Eq.~(\ref{eq:fdr1}) is preserved by this dynamics. Solving
Eq.~(\ref{eq:fpesle}) in Laplace space and putting everything
together, one recovers the conventional form of the TFR
Eq.~(\ref{eq:tfr}) for this type of dynamics. This confirms again that
a distinctive role is played by FDR1 for the validity of conventional
TFRs, even if the work PDFs are not Gaussian, as in this case.

We remark that analogous results are obtained by studying these three
types of anomalous dynamics for the case of a particle moving in a
harmonic potantial that is dragged with a constant velocity
\cite{ChKl09}.

\section{Anomalous dynamics of biological cell migration}\label{sec:cell}

In order to illustrate the application of anomalous dynamics, and
possibly of anomalous FRs, to realistic situations, in this section we
discuss experiments and theory about the migration of single
biological cells crawling on surfaces or in 3d matrices as
examples. We first introduce to the problem of cell migration by
considering cells in an equilibrium situation, i.e., not moving under
the influence of any external gradients or fields. This case is
investigated by extracting results for the mean square displacement
(MSD) and for the position PDFs from experimental data. We then show
how the experimental results can be understood by a mathematical model
in form of a fractional Klein-Kramers equation. As far as MSD and
velocity autocorrelation function are concerned, this equation bears
some similarity to a generalized Langevin equation that is of the same
type as the one that has been discussed in Section~\ref{sec:cegn}. We
finally give an outlook to the nonequilibrium problem of cell
migration under chemical gradients and describe first results obtained
from experiments and data analysis. This research paves the way to
eventually checking the existence of anomalous work TFR in biological
cell migration. The results on cell migration in equilibrium outlined
in this section are based on Ref.~\cite{DKPS08}.

\subsection{Cell migration in equilibrium}

\begin{figure}[t]  
\centerline{\includegraphics[height=10cm,angle=-90]{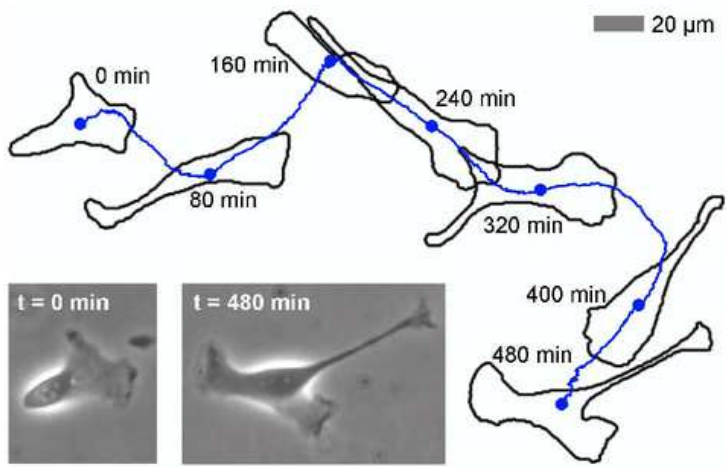}}
\caption{Overlay of a biological cell migrating {\em in vitro}
on a substrate. The cell frequently changes its shape and direction
during migration, as is shown by several cell contours extracted
during the migration process. The inset displays phase contrast images
of the cell at the beginning and to the end of its migration process
\cite{DKPS08}.}
\label{fig:cell}
\end{figure}

Nearly all cells in the human body are mobile at a given time during
their life cycle. Embryogenesis, wound-healing, immune defense and the
formation of tumor metastases are well known phenomena that rely on
cell migration \cite{LaHo96,LaSi09,FrWo10}. Fig.~\ref{fig:cell}
depicts the path of a single biological cell crawling on a substrate
measured in an {\em in vitro} experiment \cite{DKPS08}. At first
sight, the path looks like the trajectory of a Brownian particle
generated, e.g., by the ordinary Langevin dynamics of
Eq.~(\ref{eq:ole}). On the other hand, according to Einstein's theory
of Brownian motion a Brownian particle is {\em passively} driven by
collisions from the surrounding fluid molecules, whereas biological
cells move {\em actively} by themselves converting chemical into
kinetic energy. This raises the question whether the random-looking
paths of crawling biological cells can really be understood in terms
of simple Brownian motion
\cite{DuBr87,SLW91} or whether more advanced concepts of dynamical
modeling have to be applied \cite{HLCC94,URGS01,LNC08,TSYU08,BBFB10}.

\subsubsection{Experimental results}

The cell migration experiments that we now discuss have been performed
on two types of tumor-like migrating {\em transformed renal
epithelial Madin Darby canine kidney (MDCK-F)} cell strains: wild-type
($NHE^+$) and NHE-deficient ($NHE^-$) cells. Here $NHE^+$ stands for a
molecular sodium hydrogen exchanger that either is present or
deficient. It can thus be checked whether this microscopic exchanger
has an influence on cell migration, which is a typical question asked
particularly by cell physiologists. The cell diameter is about
20-50$\mu$m and the mean velocity of the cells about
$1\mu$m/min. Cells are driven by active protrusions of growing
actin filaments ({\em lamellipodial dynamics}) and coordinated
interactions with myosin motors and dynamically re-organizing
cell-substrate contacts. The leading edge dynamics of a polarized
cell proceeds at the order of seconds. Thirteen cells were observed
for up to 1000 minutes. Sequences of microscopic phase contrast images
were taken and segmented to obtain the cell boundaries shown in
Fig.~\ref{fig:cell}; see Ref.~\cite{DKPS08} for full details of the
experiments.

\begin{figure}[t]  
\centerline{\includegraphics[height=10cm,angle=-90]{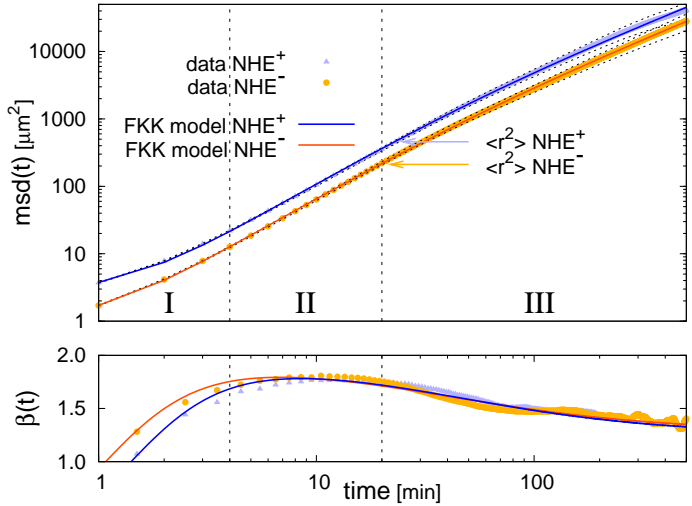}}
\caption{Upper part: Double-logarithmic plot of the mean square 
displacement (MSD) as a function of time. Experimental data points for
both cell types are shown by symbols. Different time scales are marked
as phases I, II and III as discussed in the text. The solid lines
represent fits to the MSD from the solution of our model, see
Eq.~(\ref{eq:msdn}). All parameter values of the model are given in
\cite{DKPS08}. The dashed lines indicate the uncertainties of the MSD
values according to Bayes data analysis. Lower part: Logarithmic
derivative $\beta(t)$ of the MSD for both cell types as defined by
Eq.~(\ref{eq:lder}).}
\label{fig:cell_msd}
\end{figure}

According to the Langevin description of Brownian motion outlined in
Section~\ref{sec:tfrole}, Brownian motion is characterized by a MSD
$\sigma_{x,0}^2(t)\sim t\:(t\to\infty)$ designating normal
diffusion. Fig.~\ref{fig:cell_msd} shows that both types of cells
behave differently: First of all, MDCK-F $NHE^-$ cells move less
efficiently than $NHE^+$ cells resulting in a reduced MSD for all
times. As is displayed in the upper part of this figure, the MSD of
both cell types exhibits a crossover between three different dynamical
regimes. These three phases can be best identified by extracting the
time-dependent exponent $\beta$ of the MSD $\sigma_{x,0}^2(t)\sim
t^{\beta}$ from the data, which can be done by using the logarithmic
derivative \be \beta(t)=\frac{d\ln msd(t)}{d \ln t} \quad
. \label{eq:lder} \ee The results are shown in the lower part of
Fig.~\ref{fig:cell_msd}. Phase I is characterized by an exponent
$\beta(t)$ roughly below $1.8$. In the subsequent intermediate phase
II, the MSD reaches its strongest increase with a maximum exponent
$\beta$. When the cell has approximately moved beyond a square
distance larger than its own mean square radius (indicated by arrows
in the figure), $\beta(t)$ gradually decreases to about $1.4$. Both
cell types therefore do not exhibit normal diffusion, which would be
characterized by $\beta(t)\to 1$ in the long time limit, but move
anomalously, where the exponent $\beta>1$ indicates superdiffusion.

\begin{figure}[t]
\centerline{\hspace*{-0.5cm}\includegraphics[height=9cm]{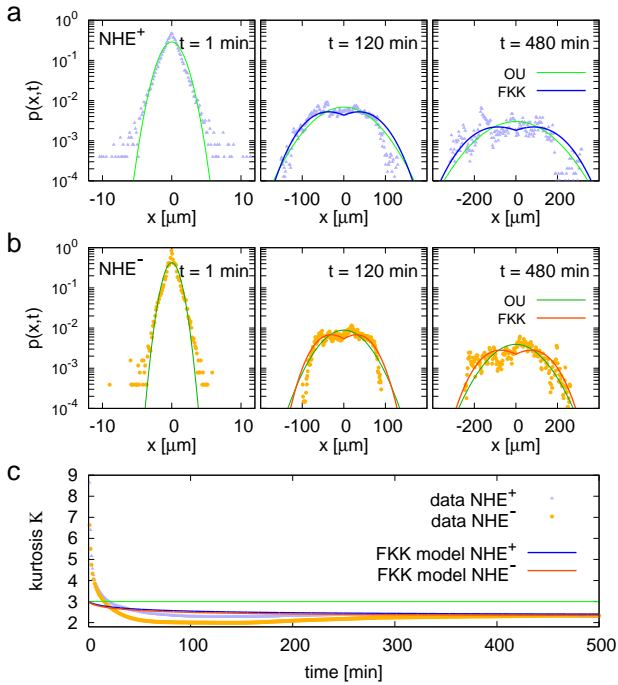}}
\caption{Spatio-temporal probability distributions $P(x,t)$. (a),(b):
Experimental data for both cell types at different times in
semilogarithmic representation. The dark lines, labeled FKK, show the
long-time asymptotic solutions of our model Eq.~(\ref{eq:fkk}) with
the same parameter set used for the MSD fit. The light lines, labeled
OU, depict fits by the Gaussian distributions Eq.~(\ref{eq:gauss})
representing Brownian motion. For $t=1$ min both $P(x,t)$ show a
peaked structure clearly deviating from a Gaussian form. (c) The
kurtosis $\kappa(t)$ of $P(x,t)$, cf.\ Eq.~(\ref{eq:kurt}), plotted as
a function of time saturates at a value different from the one of
Brownian motion (line at $\kappa=3$). The other two lines represent
$\kappa(t)$ obtained from the model Eq.~(\ref{eq:fkk}) \cite{DKPS08}.}
\label{fig:cell_pdf} \end{figure}

We next study the PDF of cell positions. Since no correlations between
$x$ and $y$ positions could be found, it suffices to restrict
ourselves to one dimension. Fig.~\ref{fig:cell_pdf} (a), (b) reveals
the existence of non-Gaussian distributions at different times. The
transition from a peaked distribution at short times to rather broad
distributions at long times suggests again the existence of distinct
dynamical processes acting on different time scales. The shape of
these distributions can be quantified by calculating the {\em
kurtosis}
\be
\kappa(t):=\frac{\langle x^4(t)\rangle}{\langle x^2(t)\rangle^2} \quad ,\label{eq:kurt}
\ee
which is displayed as a function of time in Fig.~\ref{fig:cell_pdf}
(c). For both cell types $\kappa(t)$ rapidly decays to a constant that
is clearly below three in the long time limit. A value of three would
be the result for the spreading Gaussian distributions characterizing
Brownian motion. These findings are another strong manifestation of
the anomalous nature of cell migration.

\subsubsection{Theoretical modeling}

We now present the stochastic model that we have used to reproduce the
experimental data yielding the fit functions shown in the previous two
figures. The model is defined by the {\em fractional Klein-Kramers
equation} \cite{BaSi00} \be
\pard{\varrho}{t}=-\pard{}{x}\left[v\varrho\right]+
\pard{^{1-\alpha}}{t^{1-\alpha}}\gamma_{\alpha}\left[\pard{}{v}v+
v_{th}^2\pard{^2}{v^2}\right]\varrho \quad
\:,\:0<\alpha<1\:. \label{eq:fkk}
\ee 
Here $\varrho=\varrho(x,v,t)$ is the PDF depending on time $t$,
position $x$ and velocity $v$ in one dimension, $\gamma_{\alpha}$ is a
friction term and $v_{th}^2=k_B T/M$ stands for the thermal velocity
squared of a particle of mass $M=1$ at temperature $T$, where $k_B$ is
Boltzmann's constant. The last term in this equation models diffusion
in velocity space. In contrast to Fokker-Planck equations such as
Eq.~(\ref{eq:sffpe}), this equation features time evolution both in
position and velocity space. What distinguishes this equation from an
ordinary Klein-Kramers equation, the most general model of Brownian
motion \cite{Risk}, is the presence of the Riemann-Liouville
fractional derivative of order $1-\alpha$, Eq.~(\ref{eq:rlfd}), in
front of the terms in square brackets. Note that for $\alpha=1$ the
ordinary Klein-Kramers equation is recovered. The analytical solution
of this equation for the MSD has been calculated in Ref.~\cite{BaSi00}
to
\be
\sigma_{x,0}^2(t)=2v_{th}^2t^2E_{\alpha,3}(-\gamma_{\alpha} t^{\alpha})
\quad \to\quad 2\frac{D_{\alpha}t^{2-\alpha}}{\Gamma(3-\alpha)}\quad
(t\to\infty) \label{eq:fkkmsd} 
\ee 
with $D_{\alpha}=v_{th}^2/\gamma_{\alpha}$ and the {\em
two-parametric} or {\em generalized Mittag-Leffler function} (see,
e.g., Chapter 4 of Ref.~\cite{KRS08} and Refs.~\cite{GoMa97,Podl99})
\be
E_{\alpha,\beta}(z)=\sum_{k=0}^{\infty}\frac{z^k}{\Gamma(\alpha
k+\beta)}\:,\:\alpha\:,\:\beta>0\:,\:z\in\mathbb{C} \quad .
\ee 
Note that $E_{1,1}(z)=\exp(z)$, hence $E_{\alpha,\beta}(z)$ is a
generalized exponential function. We see that for long times
Eq.~(\ref{eq:fkkmsd}) yields a power law, which reduces to the
long-time Brownian motion result in case of $\alpha=1$. 

In view of the experimental data shown in Fig.~\ref{fig:cell_msd},
Eq.~(\ref{eq:fkkmsd}) was amended by including the impact of random
perturbations acting on very short time scales for which we take
Gaussian white noise of variance $\eta^2$. This leads to \cite{MFK02}
\be
\sigma_{x,0;noise}^2(t)=\sigma_{x,0}^2(t)+2\eta^2 \quad
. \label{eq:msdn} 
\ee 
The second term mimicks both measurement errors and fluctuations of
the cell cytoskeleton. In case of the experiments with MDCK-F
cells \cite{DKPS08}, the value of $\eta$ can be extracted from the
experimental data and is larger than the estimated measurement
error. Hence, this noise must largely be of a biological nature and
may be understood as being generated by microscopic fluctuations of
the lamellipodia in the experiment.

The analytical solution of Eq.~(\ref{eq:fkk}) for $\varrho(x,v,t)$ is
not known, however, for large friction $\gamma_{\alpha}$ this equation
boils down to a fractional diffusion equation for which $\varrho(x,t)$
can be calculated in terms of a Fox function \cite{SchnWy89}. The
experimental data in Figs.~\ref{fig:cell_msd} and \ref{fig:cell_pdf}
was then fitted consistently by using the above solutions with the
four parameters $v_{th}^2, \alpha,\gamma$ and $\eta^2$ in Bayesian
data analysis \cite{DKPS08}.

In summary, by statistical analysis of experimental data we have shown
that the equilibrium migration of the biological cells under
consideration is anomalous. Related anomalies have also been observed
for other types of migrating cells
\cite{HLCC94,URGS01,LNC08,TSYU08,BBFB10}.  These experimental results
are coherently reproduced by a mathematical model in form of a
stochastic fractional equation. We now elaborate on possible physical
and biological interpretations of our findings.

First of all, we remark that the solutions of Eq.~(\ref{eq:fkk}) for
both the MSD and the velocity autocorrelation function match precisely
to the solutions of the generalized Langevin equation \cite{Lutz01}
\be 
\dot{v}=-\int_0^tdt'\:\gamma(t-t')
v(t')+\:\xi(t) \label{eq:ugle} \:.
\ee 
Here $\xi(t)$ holds for Gaussian white noise and $\gamma(t)\sim
t^{-\alpha}$ for a time-dependent friction coefficient with a power
law memory kernel, which alternatively could be written by using a
fractional derivative \cite{Lutz01}. For $\gamma(t)\sim\delta(t)$ the
ordinary Langevin equation is recovered. Note that the position PDF
generated by this equation is Gaussian in the long time limit and thus
does not match to the one of the fractional Klein-Kramers equation
Eq.~(\ref{eq:fkk}). However, alternatively one could sample from a
non-Gaussian $\xi(t)$ to generate a non-Gaussian position
PDF. Strictly speaking, despite equivalent MSD and velocity
correlations Eqs.~(\ref{eq:fkk}) and (\ref{eq:ugle}) define different
classes of anomalous stochastic processes. The precise cross-links
between the Langevin description and the fractional Klein-Kramers
equation are subtle \cite{EFJK07} and to some extent still
unknown. The advantage of Eq.~(\ref{eq:ugle}) is that it allows more
straightforwardly a possible biophysical interpretation of the origin
of the observed anomalous MSD and velocity correlations, at least
partially, in terms of the existence of a memory-dependent friction
coefficient. The latter, in turn, might be explained by anomalous
rheological properties of the cell cytoskeleton, which consists of a
complex biopolymer gel
\cite{SSGMBK07}.

Secondly, what could be the possible biological significance of the
observed anomalous cell migration? There is an ongoing debate about
whether biological organisms such as, e.g., albatrosses, marine
predators and fruit flies have managed to mimimize the search time for
food in a way that matches to optimizing search strategies in terms of
stochastic processes; see Refs.~\cite{BLMV11,VLRS11} and further
references therein. In particular, it has been argued that L\'evy
flights are superior to Brownian motion in order to find sparsely,
randomly distributed, replenishing food sources
\cite{BLMV11}. However, it was also shown that in other situations
{\em intermittent dynamics} is more efficient than pure L\'evy motion
\cite{BLMV11}.  For our cell experiment, both the experimental data
and the theoretical modeling suggest that there exists a slow
diffusion on short time scales, whereas the long-time motion is much
faster, which resembles intermittency as discussed in
Ref.~\cite{BLMV11}.  Hence, the results on anomalous cell migration
presented above might be biologically relevant in view of suitably
optimized foraging strategies.

\subsection{Cell migration under chemical gradients}

\begin{figure}[t]
\centerline{\includegraphics[width=12cm]{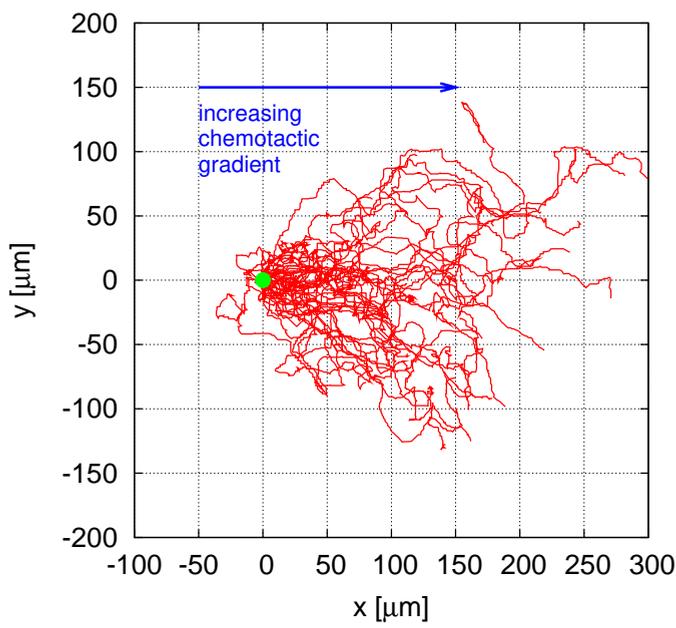}}
\caption{Trajectories of an ensemble of $40$ murine neutrophil 
cells exposed to a chemical gradient of the chemoattractant fMLP which
increases along the positive $x$-axis. Cells were observed over $30$
min with a time interval of $5$ sec. Starting points have been
transformed to the origin of the coordinate system (filled circle). It
can be seen that there is an average drift of the ensemble towards the
positive $x$-axis \cite{LS11}.}
\label{fig:cellgrad}
\end{figure}

We conclude this section with a brief outlook to cell migration under
chemical gradients \cite{Song06}. In new experiments conducted by
Lindemann and Schwab \cite{LS11}, {\em murine neutrophil} cells have
been exposed to concentration gradients of chemo-attractants. A plot
of trajectories of an ensemble of cells crawling under chemotaxis is
shown in Fig.~\ref{fig:cellgrad}.

Statistical analysis of the experimental data \cite{Diet11} yielded a
linear drift in the direction of the gradient,
\be
\langle x(t)\rangle\sim t\quad .
\ee
The MSD in the co-moving frame, on the other hand, was found to be
\be
\sigma_{x,F}^2(t)-\langle x(t)\rangle^2\sim t^{\beta}
\ee
with the same exponent $\beta>1$ as obtained for the equilibrium
dynamics discussed before. Consequently FDR1, cf.\
Eq.~(\ref{eq:fdr1}), is broken. These results suggest that, for
obtaining a stochastic model, the force-free fractional
Klein-Kramers equation Eq.~(\ref{eq:fkk}) needs to be generalized by
including an external force as discussed by Metzler and Sokolov
\cite{MeSo02},
\be
\pard{\varrho}{t}=-\pard{}{x}\left[v\varrho\right]+\pard{^{1-\alpha}}{t^{1-\alpha}}\gamma_{\alpha}\left[\pard{}{v}v-\frac{F}{\gamma_{\alpha} m}\pard{}{v}+v_{th}^2\pard{^2}{v^2}\right]\varrho \label{eq:fkkneq}\:.
\ee
Note that there exist two different ways in the literature of how to
include the force $F$ in Eq.~(\ref{eq:fkk}) \cite{BaSi00,MeSo02}.
These choices lead to different results for drift and MSD. The above
results obtained from experimental data analysis clearly select the
version of Ref.~\cite{MeSo02} as the adequate type of stochastic
model in this case by rejecting that of Ref.~\cite{BaSi00} which,
however, might well work in other situations. The (approximate)
analytical solutions of Eq.~(\ref{eq:fkkneq}) reproduce correctly
drift, MSD, velocity correlations and (for large enough friction
coefficient $\gamma_{\alpha}$ and long enough times) the position PDFs
of the measured nonequilibrium cell dynamics \cite{Diet11}.

\begin{figure}[t]
\centerline{\includegraphics[width=12cm,angle=0]{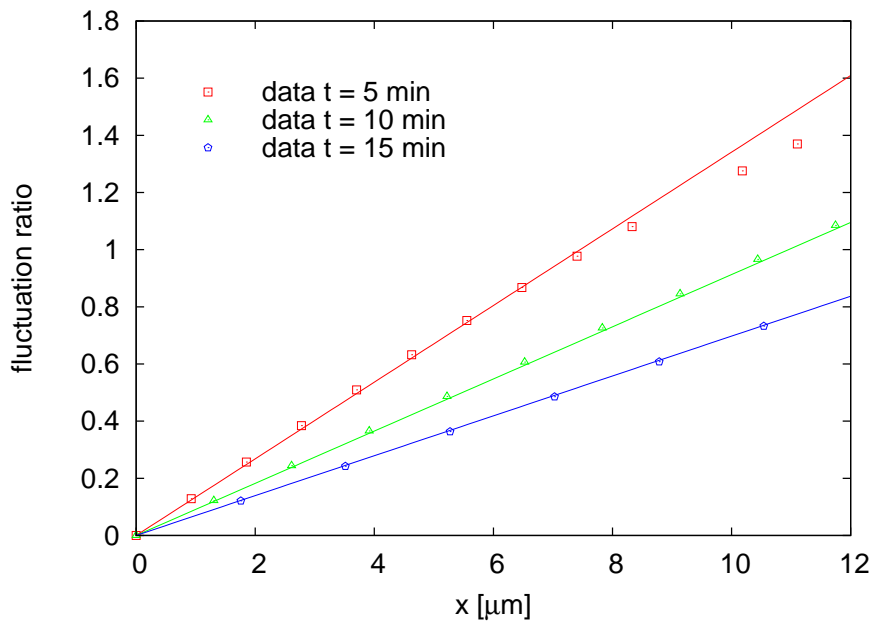}}
\caption{The fluctuation ratio $\ln(P(A)/P(-A))$  as a function of
$A=x(\tau)-x(0)$ for $\tau=5$ min (open squares), $\tau=10$ min (open
triangles) and $\tau=15$ min (open pentagons) obtained from 90
independent trajectories of murine neutrophils moving in a chemotactic
gradient of the substance fMLP as depicted in Fig.~\ref{fig:cellgrad}.
Data show a linear increase in $A$, however, the reduction of the
slope as a function of $\tau$ indicates deviations from the
conventional TFR Eq.~(\ref{eq:tfr}).}
\label{fig:cellafr}
\end{figure}

Along these lines, one might also check for the form of the work TFR
in case of cell migration. This has already been done in an experiment
on the cellular slime mold {\em Dictyostelium discoideum}, in this
case under electrotaxis \cite{HaTa07}. By plotting the fluctuation
ratio as a function of the cell positions at two different times it
was concluded that the conventional TFR Eq.~(\ref{eq:tfr}) was
confirmed by this experiment. In Fig.~\ref{fig:cellafr}, however, we
show experimental results for the fluctuation ratio of the neutrophils
of Fig.~\ref{fig:cellgrad} as a function of the cell positions at
three different times. In complete formal analogy to
Fig.~\ref{fig:sellafr}, the slopes clearly decrease with increasing
time, which indicates a violation of the conventional TFR
Eq.~(\ref{eq:tfr}). To further explore the validity of work TFRs in
cell migration experiments thus appears to be a very interesting,
important open problem.

\section{Conclusions}\label{sec:suco}

In this chapter we have applied the concept of FRs as discussed in the
previous book chapters to anomalous stochastic processes. This
cross-linking enables us to address the question whether conventional
forms of FRs are valid for more complicated types of dynamics
involving non-Markovian memory and non-Gaussian distributions. We have
answered this question for three fundamental types of anomalous
stochastic dynamics:

For {\em Gaussian stochastic processes with correlated noise} the
existence of FDR2 implies the existence of FDR1, and we have found
that FDR1 in turn implies the existence of work TFR in conventional
form. That is, analytical calculations showed that the conventional
work TFR holds for internal noise. However, a weak violation of the
conventional form was detected in case of external noise yielding a
pre-factor that is not equal to one and in particular depends on
time. A strong violation of the conventional work TFR was derived for
{\em space-fractional L\'evy dynamics} confirming previous results
from the literature. We have also found that the conventional work TFR
holds for a typical example of {\em time-fractional dynamics}.  These
generic models suggest an intimate connection between FDRs and FRs in
case of anomalous dynamics.

As a realistic example of anomalous dynamics, we have then discussed
biological cell migration. By extracting the MSD and the position PDF
from experimental data for cells crawling in an equilibrium {\em in
vitro} situation, we found that the cells under investigation
exhibited different dynamics on different time scales deviating from
simple Brownian motion. For longer times, these cells moved
superdiffusively. These experimental findings were reproduced by a
stochastic model in form of a fractional Klein-Kramers equation. For
cells moving in nonequilibrium under chemotaxis, new data showed a
breaking of FDR1 leading to a stochastic modeling in form of a
suitably extended fractional Klein-Kramers equation. Further analysis
of this data indicated the existence of anomalous work TFRs.

To better understand work TFRs in biological cell migration both
theoretically and experimentally remains an important open
problem. However, it might also be interesting to experimentally check
for anomalous work TFR in case of a particle dragged through a highly
viscous gel instead of through water \cite{WSM+02}, for the
fluctuations of a driven pendulum in gel
\cite{DJGPC06}, for granular gases exhibiting subdiffusion
\cite{BrPoe04}, or for glassy systems
\cite{ZRA05,ZBC05,Sell09}. On the theoretical side, the basic results
reported in this chapter suggest to systematically check the remaining
variety of conventional fluctuation relations
\cite{Croo99,HaSa01,Sei05} for anomalous generalizations.

\subsection*{Acknowledgments}

R.K.\ and A.V.C.\ wish to thank the London Mathematical Society for a
travel grant.  Financial support by the EPSRC for a small grant within
the framework of the QMUL {\em Bridging The Gap} initiative is
gratefully acknowledged.

%\bibliography{summ17}
%\bibliographystyle{wivchnum}

\end{document}